\documentstyle[preprint,prc,aps,epsfig,epsf]{revtex}
\begin{document}
\title{ON THE ORIGIN OF THE SHORT RANGE NN REPULSION}
\author{ K. Shimizu$^1$ and L. Ya. Glozman$^2$}
\address{ $^1$ Department of Physics,Sophia University, 
Tokyo 102-8554, Japan\\
$^2$ High Energy Accelerator Research Organization (KEK),
Tanashi Branch, Tanashi, Tokyo 188-8501, Japan
\footnote{e-mail: lyg@cleopatra.kfunigraz.ac.at; present 
address: Institute for Theoretical Physics,
University of Graz, Universit\"atsplatz 5, A-8010 Graz,
Austria}}
\maketitle

\begin{abstract}
We calculate S-wave singlet and triplet NN phase shifts
stemming from the short-range flavor-spin hyperfine
interaction between constituent quarks using the resonating
group method approach. A strong short-range repulsion is
found in both waves. A fair comparison is performed between
the traditional picture, relying on the colour-magnetic
interaction, and the present one, relying on the Goldstone
boson exchange dynamics. It is shown that the latter one
induces essentially stronger repulsion, which is a very
welcome feature. We also study a sensitivity of phase shifts
and wave function to extention from the one-channel
to three-channel resonating group method approximation.
\end{abstract}

\bigskip
\bigskip

One of the crucial questions of the low-energy QCD is which physics,
inherent in QCD, is responsible for the low-energy properties
of light and strange baryons and their interactions. At the
very high momenta transfer (the ultraviolet regime of QCD)
the nucleon is viewed as a system of the weakly interacting partons,
which justifies here the use of the perturbative QCD tools.
The low-energy properties, like masses or low-energy interactions,
is much harder, if impossible, to understand in terms of the original
QCD degrees of freedom and one obviously needs effective theories.
One can borrow a wisdom of the many fermion system physics,
e.g. in condensed matter, which suggests that in such a situation
the concept of the quasiparticles in Bogoliubov or Landau sense
becomes very useful. The idea of quasiparticles is that in some
circumstances one can approximately absorb complicated 
interactions between bare fermions, i.e. in our case current quarks,
into static properties of quasiparticles, e.g. their masses,
and what is left beyond that should be treated as residual
interactions between quasiparticles. Such a concept should be
helpful to understand the low-energy properties of light baryons 
where a typical momentum of quarks  is {\it below}
the chiral symmetry breaking scale, $\Lambda_\chi \sim 1$ GeV, which
implies that the low-energy characteristics of baryons, 
such as masses, are
formed by the nonperturbative QCD dynamics, which is responsible
for the chiral symmetry breaking and confinement, 
but not by the perturbative
QCD interactions  which should be active
at much higher scale, where the quasiparticles {\it do not exist}.  
At low momenta the scalar part of the nonperturbative gluonic interaction
between current quarks, which triggers the chiral symmetry
breaking (i.e.  pairs the left quarks and right
antiquarks and vice versa in the QCD vacuum), can be absorbed
into the mass of quasiparticles - constituent quarks. 
At the same time this nonperturbative interaction, iterated in the $qq$ t-channel in baryons, leads
to the poles which can be identified as Goldstone boson exchange
between valence quarks in baryons\cite{GV}. This is a general 
feature and does not depend on details or
nature of this nonperturbative interaction. If so, the adequate
{\it residual} interactions between the constituent quarks in baryons
{\it at low momenta}, $q < \Lambda_\chi$, should be effective confining interaction
and the Goldstone boson exchange (GBE). \\

By now it is established that such a picture is very successful
in light and strange baryon spectroscopy \cite{GR,GPVW}. 
Similar
conclusions have been obtained recently in the lattice studies
of $N-\Delta$ splitting \cite{LIU}, in the large $N_c$ \cite{CARLSON} 
and phenomenological \cite{GEORGI} analyses of L=1  spectra. 
 If such
a physical picture is satisfactory, it should also explain 
baryon-baryon interaction. It is rather evident that at medium and
large distances in the baryon-baryon system, where the Pauli
principle at the constituent quark level does not play any role,
it is fully compatible with the wisdom of nuclear physics,
where the $NN$ interaction is determined by the Yukawa tail
of the pion exchange and the 
two-pion exchange ($\rho$- and $\sigma$- exchange
interactions). The explanation that the short-range repulsion
is due to the central spin-independent part of the $\omega$ exchange
is not satisfactory, however, as in this case the $\omega N$
coupling constant should be increased by a factor 3  compared to
its empirical value. One takes it for granted that the origin
of the short-range $NN$ repulsion should be the same as the
origin of the nucleon mass and its lowest excitations. If so, the
Fermi nature of constituent quarks and specific interactions
between them should be of crucial importance to understand
the short range $NN$ repulsion.\\

Traditionally the repulsive core in the $NN$ system within the
constituent quark picture was attributed to the colour-magnetic
part of the one gluon exchange (OGE) interaction combined with
 quark interchanges between $3Q$ clusters (for reviews and earlier
references see \cite{OY,SH}). However,
as it follows from the previous discussion it is dubious to use
a language of constituent quarks and at the same time of perturbative
one gluon exchange. So the important question is whether one can
understand the short-range $NN$ repulsion in terms of residual
interactions  like GBE.\\

The first simple analysis of  possible effects for the 
S-wave $NN$ system 
from the short-range part of the pion-exchange interaction
between quarks was based on the assumption that the $6Q$ wave
function in the nucleon overlap region has a flavor-spin
symmetry $[33]_{FS}$, which is the only possible symmetry in
the nonexcited $s^6$ configuration \cite{SHIM}. 
In that paper, as well as in the subsequent hybrid models
\cite{KUS,VAL,ZH,FUJ}, it was assumed that the pion-exchange
produces only some insignificant part 
 of the short-range $NN$ repulsion.
However, when
the GBE-like hyperfine
interaction {\it is made strong enough to produce the
$\Delta-N$ splitting and describe the low-lying spectrum}, 
the situation is different.
The GBE-like interaction is 
more attractive within the $6Q$ configuration
with the symmetry $[51]_{FS}$ and thus the 
spatially excited configuration 
$s^4p^2 [51]_{FS}$ is more favourable and becomes the lowest
one \cite{STPGL,STGL}. The energy of this configuration is, however, 
still much higher
than the energy of two infinitely separated nucleons and that is
why there appears a strong short-range repulsion in the $NN$ system.
While this result demonstrates that the GBE-like hyperfine interaction
could indeed explain the short-range repulsive core in the NN system, it is still
only suggestive (the phase shifts have not been calculated) as it is 
based on the adiabatic approximation
and  neglects a smooth transition to the distances with
the well-clustered $6Q$ system. In the present work we go beyond
the adiabatic approximation and construct our basis in such a way
that it includes not only the lowest important $s^4p^2$ and $s^6$
configurations like in \cite{STPGL,STGL}, but also the well clustered
states at medium and long distances. We calculate both the $^3S_1$ and
$^1S_0$ phase shifts and prove that the GBE-like flavor-spin hyperfine
interaction does supply a very strong short-range repulsion in
the $NN$ system. We compare this repulsion with the one induced by
the colour-magnetic interaction  within the traditional picture
and find the former one to be much stronger.
This is a very welcome feature as the models of the short-range
$NN$ repulsion based on the OGE
interaction \cite{OY,SH,KUS,VAL,ZH,FUJ} fail to describe phase
shifts above the lab energy of about 300 MeV because of the lack
of the strong enough short-range repulsion in those models.\\

The convenient basis for solving the Schr\"odinger equation which
comprises both the short-range $6Q$ configurations, incorporates the
$NN$ asymptotics as well as a smooth transition from the nucleon
overlap region to the medium ranges is suggested by the resonating
group method (RGM) approximation. The most simple one-channel
ansatz for the six-quark two-nucleon wave function is

\begin{eqnarray}
\psi & = & \hat{A}\{N(1,2,3)N(4,5,6)\chi(\vec{r})\}, \label{WFI}\\
\nonumber
\hat{A}&=& \frac{1}{\sqrt{10}}(1-9 \hat{P}_{36}), \\
\nonumber
\vec{r}&=& \frac{\vec{r}_1+\vec{r}_2+\vec{r}_3}{3} -
\frac{\vec{r}_4+\vec{r}_5+\vec{r}_6}{3}.
\end{eqnarray}

\noindent
Here $N(1,2,3)$ is $s^3$ harmonic oscillator wave function of the
nucleon with a standard $SU(6)_{FS}$ spin-isospin part, $\hat{A}$
is an antisymmetrizer at the quark level and the trial function 
$\chi(\vec{r})$ is obtained by solving the Schr\"odinger equation,
for review see \cite{OY,SH}. We remind, however, that
at short range the trial
function $\chi(\vec{r})$ by no means should be interpreted as a
relative motion wave function: in the nucleon overlap region
because of the antisymmetrizer
the function (\ref{WFI}) is intrinsically 6-body function and
contains a lot of other ``baryon-baryon components'', such as
$\Delta \Delta$, $NN^*$, $N^*N^*$,... and hidden colour components
\cite{GLKU}.

 The trial function (\ref{WFI}) completely includes the 
 symmetric short-range
$s^6$ shell-model configuration provided that the harmonic oscillator
parameter for the $s^3$ nucleon and for $s^6$ configuration
coinside. This is because of the well-known identity

\begin{equation}
\hat{A}\{N(1,2,3)N(4,5,6)\phi_{0s}(\vec{r})\}_{SI} =
\sqrt{\frac {10}{9}} |s^6[6]_O [33]_{FS}>. \label{s6} 
\end{equation}

\noindent
Here and below $\phi_{Ns}(\vec {r})$ denotes the S-wave
harmonic oscillator function with $N$ excitation quanta,
$[f]_O$ and $[f]_{FS}$ are Young diagrams (patterns)
describing the permutational orbital and flavour-spin
symmetries in $6Q$ system,
which are necessary to identify the given configuration in the 
shell-model basis. It is always assumed that
the center-of-mass motion is removed from the shell-model
wave function.

However, the ansatz (\ref{WFI})
contains only a fixed superposition of different shell-model
configurations from the $s^4p^2$ shell \cite{STPGL}:

\begin{eqnarray}
\hat{A}\{N(1,2,3)N(4,5,6)\phi_{2s}(\vec{r})\}_{SI} \nonumber \\
= \frac {3\sqrt{2}}{9}|(\sqrt{\frac {5}{6}} s^52s 
 - \sqrt{\frac {1}{6}} s^4p^2) [6]_O [33]_{FS}> \nonumber \\
-\frac {4 \sqrt{2}}{9} | s^4p^2 [42]_O [33]_{FS}> \nonumber \\
 -\frac {4 \sqrt{2}}{9} | s^4p^2 [42]_O
[51]_{FS}>. \label{s4p2}
\end{eqnarray}

One can extend the ansatz (\ref{WFI}) and include in addition
two new channels, ``the $\Delta \Delta$'' and the ``hidden colour
channel CC'' \cite{HARVEY,FFLS}:

\begin{eqnarray}
\psi  =  \hat{A}\{N(1,2,3)N(4,5,6)\chi_{NN}(\vec{r})\} \nonumber \\
+  \hat{A}\{\Delta(1,2,3)\Delta(4,5,6)\chi_{\Delta\Delta}(\vec{r})\}
\nonumber \\
+  \hat{A}\{C(1,2,3)C(4,5,6)\chi_{CC}(\vec{r})\}
, \label{WFIII}
\end{eqnarray}

\noindent
where $\Delta(1,2,3)$ is $s^3$ harmonic oscillator $SU(6)_{FS}$
wave function of the $\Delta$-resonance and the hidden-colour
$CC$ channel includes the $C=$ colour-octet
$s^3$ cluster. 
Here we followed a definition of the hidden-color channel $CC$ 
in ref. \cite{FFLS}. The hidden-color state  
 is constructed so that it contains only the flavor-spin
$[33]_{FS}$ symmetry. It must be noted that $C$ is the color 
 octet but it does not have a definite spin and isospin. 
Note that all three channels in
(\ref{WFIII}) are highly non-orthogonal because of the antisymmetrizer.
This can be easily seen from the fact that identities similar to 
(\ref{s6}) can be written also for the $\Delta \Delta$ and
$CC$ channels. This redundancy in the subspace $N=0$ is only a technical
one and can be easily avoided by diagonalizing the norm RGM
matrix and removing all ``forbidden states''.
 However, in the subspace with $N=2$, these
three channels become linearly independent since the following
identities  are also
valid for the $\Delta \Delta$ and $CC$ channels:

\begin{eqnarray}
\hat{A}\{\Delta(1,2,3)\Delta(4,5,6)\phi_{2s}(\vec{r})\}_{SI} = \nonumber \\
-\frac {6\sqrt{10}}{45}
|(\sqrt{\frac {5}{6}} s^52s - \sqrt{\frac {1}{6}} s^4p^2) [6]_O [33]_{FS}>
  \nonumber\\
+\frac {8 \sqrt{10}}{45} | s^4p^2 [42]_O [33]_{FS}> \nonumber \\ -
\frac {2 \sqrt{10}}{9} | s^4p^2 [42]_O
[51]_{FS}>, \label{DD}
\end{eqnarray}

\begin{eqnarray}
\hat{A}\{C(1,2,3)C(4,5,6)\phi_{2s}(\vec{r})\}_{SI} \nonumber \\
=
\frac {2\sqrt{10}}{5}
|(\sqrt{\frac {5}{6}} s^52s - \sqrt{\frac {1}{6}} s^4p^2) [6]_O [33]_{FS}>
  \nonumber\\
+\frac {2 \sqrt{10}}{15} | s^4p^2 [42]_O [33]_{FS}>. \label{CC}
\end{eqnarray}

\noindent
Because the trial functions $\chi_{NN}, \chi_{\Delta\Delta}$
and $\chi_{CC}$ are independent full Hilbert space trial functions (i.e.
they completely include $\phi_{2s}$), then the compact shell-model
configurations 
$|\sqrt{\frac {5}{6}}s^52s-\sqrt{\frac{1}{6}}s^4p^2[6]_O [33]_{FS}>$,
$ | s^4p^2 [42]_O [33]_{FS}>$ and $ | s^4p^2 [42]_O [51]_{FS}>$ 
are relaxed and participate as independent variational
configurations when one applies the ansatz (\ref{WFIII}),
in contrast to the ansatz (\ref{WFI}). 
The other possible compact $6Q$ configurations
from the $s^4p^2$ shell, such as  $[411]_{FS}$, $[321]_{FS}$
and $[2211]_{FS}$ are not taken into account,
 but they play only a very modest role when
one applies the interaction (\ref{opFS}) \cite{STPGL,STGL}.\\

First we  study the effect of the GBE-like flavor-spin short
range interaction. This interaction can be
parametrized as

\begin{equation}
 V_{\chi} = - \sum_{i<j} \frac{a_{\chi}}{m_im_j} \vec{\tau}_{i} 
\cdot \vec{\tau}_j
\vec{\sigma}_i \cdot \vec{\sigma}_j \Lambda^2 \frac{e^{-\Lambda r}}{r} ,
\label{opFS}
\end{equation}

\noindent
where $\vec{\tau}$
and $\vec{\sigma}$ are the quark isospin
 and spin matrices respectively.
In this qualitative paper we confine ourselves to the $\pi$-exchange between
$u,d$ quarks as the contribution of the $\eta$ exchange is much
smaller. 

The minus sign of the interaction (\ref{opFS}) is
related to the sign of the short-range part of the pseudoscalar
 meson-exchange interaction (which is opposite to that 
of the Yukawa tail),
crucial for the hyperfine splittings in baryon spectroscopy.
It is significant that this short-range part appears at
the leading order within the chiral perturbation theory
(i.e. in the chiral limit), while the Yukawa part contributes
only in the subleading orders and vanishes in the chiral limit
\cite{G}.
The parameter $\Lambda$, which determines a range of this interaction,
 is fixed by the scale of spontaneous
breaking of chiral symmetry, $\Lambda \simeq 1$ GeV. 
The Yukawa part of the interaction, on the other hand, is determined
by the pion mass and is not important for the interaction of
quarks at distances of $0.5 - 0.8$ fm, which is a typical distance
between quarks in the nucleon and which is important for the
short-range $NN$ interaction. Note that the short-range
interaction of the form (\ref{opFS}) comes also from the
$\rho$-exchange \cite{GLOZ}, which can also be considered
as a representation of the correlated two-pion exchange \cite{RB}.
The parameter
$a_{\chi}$, which determines the total strength of the
pseudoscalar and vector-like hyperfine interactions
with $s^3$ ansatz for both nucleon and $\Delta$ wave function 
is fixed to
reproduce the $\Delta - N$ mass splitting. 
 The constituent
masses are taken to have their typical values, $m=\frac{1}{3}m_N$.
When the confining interaction between quarks is assumed to be
colour-electric and pairwise and has a harmonic form, it does
not contribute at all to the two nucleon problem as soon as
the ans\"atze (\ref{WFI}) or (\ref{WFIII}) are 
used and the two-nucleon threshold,
calculated with the same Hamiltonian, is subtracted. Hence 
within the given toy model we have only one free parameter,
the nucleon matter root-mean-square size $b$, which coinsides with
the harmonic oscillator parameter of the $s^3$ wave function.
We fix it to be $b=0.5$ fm.  The parameters used in the calculation are 
summarized in Table \ref{tab1}.\\

The second model is a traditional one, based on the colour-magnetic
component of OGE

\begin{equation}
 V_{cm} = - \sum_{i<j} \frac{a_{cm}}{m_im_j} \lambda_{i}^{C} 
\cdot \lambda_{j}^{C}
\vec{\sigma}_i \cdot \vec{\sigma}_j \Lambda^2 \frac{e^{-\Lambda r}}{r},
\label{opCM}
\end{equation}

\noindent
where $\lambda^C$ are color Gell-Mann matrices with an implied
summation over $C = 1,...,8$.
We want to make a fair comparison between two models and thus use
exactly the same $b$ and $\Lambda$. The effective OGE coupling constant
$a_{cm}$ is determined from the $\Delta -N$ mass splitting,
 which is also given in the Table.  
 Thus we can study a difference between repulsion
implied by the flavor-spin and color-spin structures of the
hyperfine interactions.\\

\begin{figure}[htb]
\caption{Phase shifts for the NN $^3S_1$ channel.
Phase shifts given by the single channel calculation are
 shown for the models GBE and OGE as a function of
 the wave number $k$.} 

\centerline{\epsfig{file=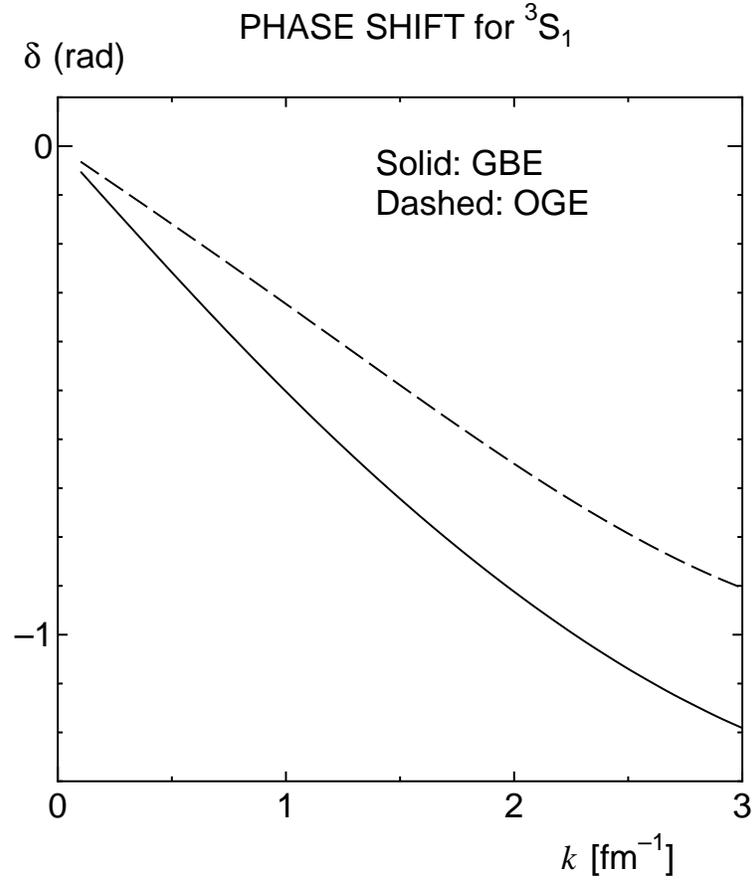,width=10cm}}
\label{fig1a}
\end{figure}

\begin{figure}[htb]
\caption{Phase shifts for the $NN { }^1S_0$ channel}
{See Fig. \ref{fig1a} for explanation}

\centerline{\epsfig{file=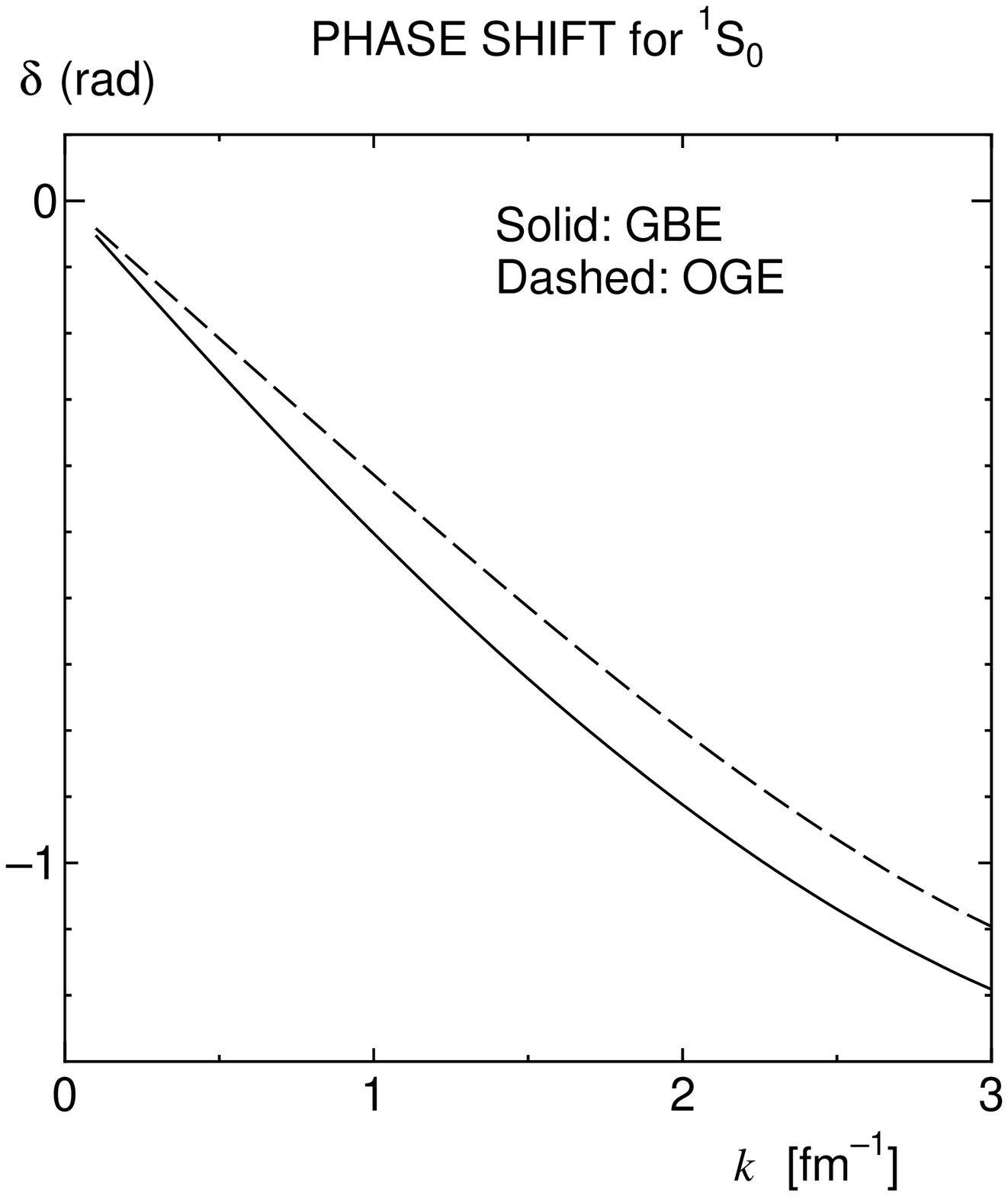,width=10cm}}
\label{fig1b}
\end{figure}

In Figures \ref{fig1a} and \ref{fig1b}, we show the S-wave triplet and 
singlet phase shifts as a function of the center of mass momentum
for both models, which are negative and thus indicate repulsion
in both cases.  However, it is immediately seen
from  comparison that the flavour-spin hyperfine interaction (\ref{opFS})
supplies essentially stronger repulsion than the colour-magnetic interaction (\ref{opCM}).
One of the reasons is that while the colour-magnetic interaction
contributes to the short-range repulsion exclusively via
the quark-exchange terms which vanish when one approaches
the nucleon size to zero, the repulsion in the NN system
stemming from the flavor-spin interaction
 is supported by both direct and quark-exchange terms and does
not vanish in this limit.\\

Next we address the issue whether an extention from the
one-channel ansatz (\ref{WFI}) to the three-channel ansatz
(\ref{WFIII}) is important for phase shifts and six-quark
wave function. We employ the model (\ref{opFS}).

\begin{figure}[htb]
\caption{Phase shifts for the NN $^3S_1$ channel.
Phase shifts given by the single and three channel
 calculations are shown for the 
models GBE and OGE interaction as a function of the wave 
number $k$. Two curves correspond to single channel and three channel 
(weaker repulsion) calculations.}

\centerline{\epsfig{file=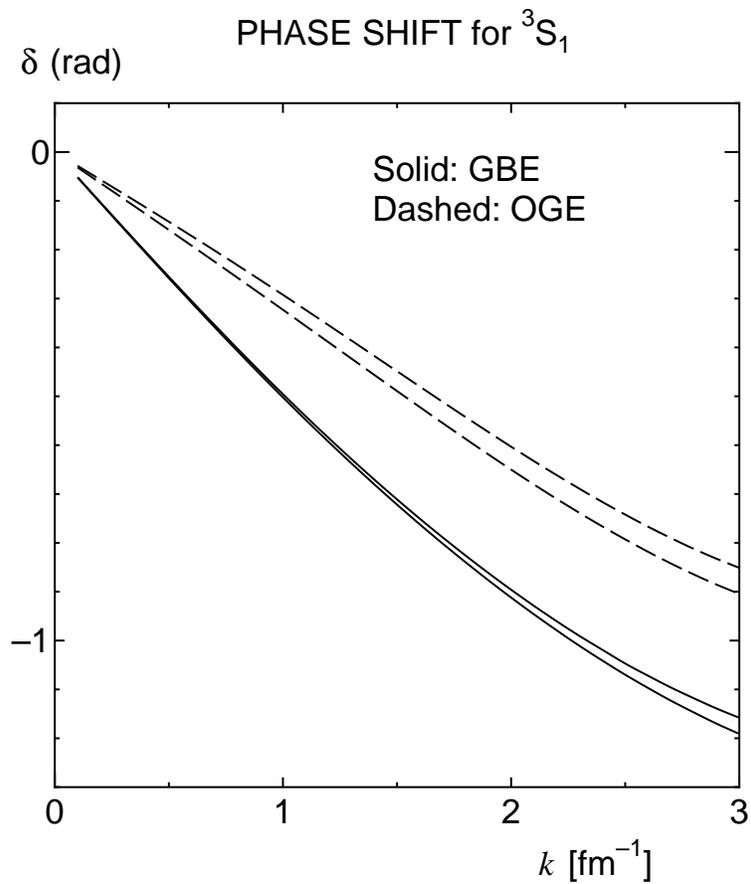,width=10cm}}
\label{fig2}
\end{figure}

In Fig. \ref{fig2} we compare phase shifts calculated with the one-channel
and the three-channel ans\"atze. While there is some difference,
it is not significant. Note that inclusion of the $\Delta\Delta$
channel will produce an important effect as soon as the
long- and intermediate-range attraction between quarks 
( $\pi$, 2$\pi$ or $\sigma$ exchanges) is included.

There is, however, a  difference
in the short-range 6Q wave functions. Unfortunately it is
not possible to show the six-body wave function in both cases,
but we can compare a projection of the wave function onto
the given baryon-baryon component. There is no unique
definition of such a projection, because it is not
an observable and does not make a direct physical sense
(for a discussion on this issue see ref. \cite{GLKU}).
Only a full 6-body wave function can be used to calculate
any observable, which includes both the direct and the quark-interchange
terms. We shall use two different definitions, one of them
via the first power of the norm kernel (this correspond
to that one used in \cite{KUS,GLKU})

\begin{equation}
{\bar \chi}_\alpha({\vec r}'') = \int d{\vec r}' N_{\beta \alpha}
({\vec r}' {\vec r}'') \chi_\beta({\vec r}'),
\label{P1}
\end{equation}

\begin{equation}
  N_{\beta \alpha} ({\vec r}', {\vec r}'') =
<B_\beta(1,2,3) B_\beta(4,5,6)\delta({\vec r} - {\vec r}')
| 1 -9 {\hat P}_{36} |
B_\alpha(1,2,3) B_\alpha(4,5,6)\delta({\vec r} - {\vec r}'')>,
\label{N}
\end{equation}

\noindent
where $B_\alpha = N, \Delta, C$. The other definition uses
a square root of the norm kernel

\begin{equation}
{\bar \chi}'_\alpha({\vec r}'') = \int d{\vec r}' N^{1/2}_{\beta \alpha}
({\vec r}' {\vec r}'') \chi_\beta({\vec r}').
\label{P2}
\end{equation}

\noindent
Sometimes the latter projection is interpreted as a probability
density for a given channel, which is, however, not correct,
since only the full 6-body wave function has a direct
and clear probability interpretation.

Both types of projections would give an identiacal result
if one used a multichannel ansatz for wave function
with all possible baryon states. Then the closure relation

\begin{equation}
\sum_\alpha | B_\alpha(1,2,3) B_\alpha(4,5,6) >
< B_\alpha(1,2,3) B_\alpha(4,5,6)| = I
\label{CLOSURE}
\end{equation}

\noindent
would be satisfied and one would obtain $ \hat N  = ({\hat N}^{1/2})^2$
(which is satisfied on the subspace $N=0$
but not satisfied on the subspace $N=2$ 
with $B_\alpha = N,\Delta,C$).

\begin{figure}[htb]
\caption{Projection of wave function on the NN $^3S_1$ channel.
The results of the single and three channel calculations using the model
GBE are shown. $N$ and $N^{1/2}$ correspond to the projections
 using the norm kernel and a square root of the norm kernel, respectively.}

\centerline{\epsfig{file=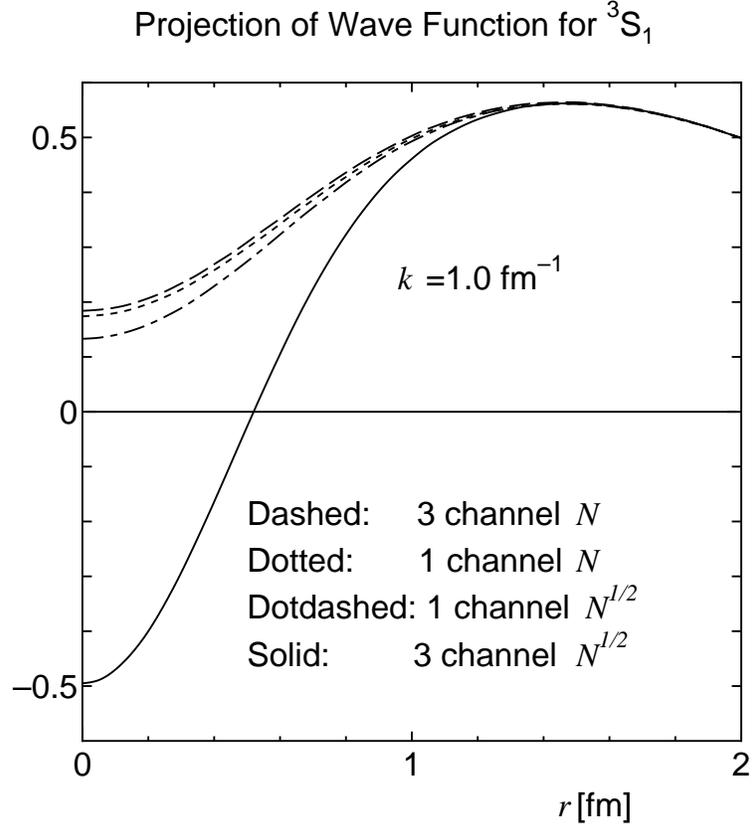,width=10cm}}
\label{fig3}
\end{figure}

In Figs. \ref{fig3} we show projections onto $NN$ 
using both one-channel and three-channel ans\"atze and both
definitions of projections. It is indeed well seen that 
different definitions give different behaviour of projections
at short range. While there is a node with the definition
(\ref{P2}), such a node is absent with the definition (\ref{P1}),
which illustrates a very limited physical sense of  
projections. Still, when we compare the projections obtained
with different ans\"atze (\ref{WFI}) or (\ref{WFIII}) within
the same definition (\ref{P2}), one observes a significant
difference, which is of no surprise since the ansatz (\ref{WFIII})
is much richer at short distances in the NN system.\\

In conclusion we summarize. The short-range flavour-spin hyperfine
interaction between the constituent quarks implies a strong short-range
repulsion in the NN system. This repulsion is essentially stronger
than that one supplied by the colour-magnetic interaction within
the traditional model. This is a welcome feature as the traditional
models, based on the colour-magnetic interaction, do not provide a strong
enough short-range repulsion and fail to
describe the phase shifts above the lab energy of 300 MeV.
Another significant difference is that the interaction (\ref{opFS})
implies a repulsion of the same strength in both singlet and
triplet partial waves, while the colour-magnetic interaction
supplies a repulsion of different strength, which makes it difficult
to describe both partial waves at the same time. While both
models imply a repulsion in the $NN$ system, their implications
are dramatically different in the "H-particle" channel. The
colour-magnetic interaction, reinforced by the the Yukawa parts
of the meson exchanges, leads to a deeply bound H-particle \cite{S},
while the interaction (\ref{opFS}) tends to make the  $6q$ system
with "H-particle" quantum numbers unbound or loosly bound\cite{SPG2}. 
The existing
experimental data exclude the deeply bound H-particle \cite{HYP97}.\\

Thus the chiral constituent quark model has
a good potential to explain not only baryon spectroscopy, but
also the baryon-baryon interaction. The next stage is to add
a long-range Yukawa potential tail from one-pion and two-pion
(sigma + rho) exchanges (and possibly from omega-exchange) and
provide a realistic description of NN system including all
the necessary spin-spin, tensor and spin-orbit components. This
task is rather involved and all groups with the corresponding
experience are invited.\\

L.Ya.G. acknowledges a warm hospitality of the nuclear
theory groups of KEK-Tanashi and Tokyo Institute of Technology.
His work is supported by the foreign guestprofessorship
program of the Ministry of Education, Science, Sports and
Culture of Japan.

\begin{table}
\caption{Parameters of models GBE and OGE interaction.
 $a_c$ is the strength of the 
harmonic confinement.}
\vspace{5mm}

\begin{tabular}{| c | r | r | r | r | r | r |} \hline
Model & $m$ [MeV] & $b$ [fm] & $\Lambda$ [GeV] &$a_{cm}$ & $a_{\chi}$
& $a_c$ [MeV/fm$^2$]  \\ \hline
GBE  & 313 & 0.5 & 1.0 & 0.0 & 0.068 & 47.7 
\\ \hline 
OGE & 313 & 0.5 & 1.0 & 0.051 & 0.0 &  93.7 \\ \hline
\end{tabular}
\label{tab1}
\end{table}

\newpage

\newpage

\newpage

\newpage

\end{document}